\documentclass[aps,pra,reprint,superscriptaddress]{revtex4-1}
\usepackage{amsmath,amsfonts,amssymb}
\usepackage{graphicx,calc,bm,physics,slashed}

\usepackage[breaklinks,colorlinks,urlcolor=blue,citecolor=blue,linkcolor=blue]{hyperref}
\usepackage{color}
\usepackage{caption}
\usepackage{setspace}
\usepackage{subcaption}
\usepackage{float}
\usepackage{tikz}
\usepackage[format=plain,justification=raggedright,singlelinecheck=false]{caption}
\begin{document} 

\captionsetup[figure]{labelformat={default},labelsep=period,name={FIG.}}
\title{Emergence of  charge-density-wave and superconduct phase transition through Lorentz-ivariant interactions in the Haldane-Hubbard model}
\author{Qiao Yang}
\thanks{These authors contributed equally to this work}
\affiliation{Beijing National Laboratory for Condensed Matter Physics, Institute of Physics, Chinese Academy of Sciences, Beijing 100190, China}
\affiliation{University of Chinese Academy of Sciences, Beijing 100049, China}
\author{Yu-Biao Wu}
\thanks{These authors contributed equally to this work}
\affiliation{Beijing National Laboratory for Condensed Matter Physics, Institute of Physics, Chinese Academy of Sciences, Beijing 100190, China}

\author{Lin Zhuang}
\affiliation{State Key Laboratory of Optoelectronic Materials and Technologies, School of Physics, Sun Yat-Sen University, Guangzhou 510275, China}

\author{Ji-Min Zhao}
\affiliation{Beijing National Laboratory for Condensed Matter Physics, Institute of Physics, Chinese Academy of Sciences, Beijing 100190, China}
\affiliation{University of Chinese Academy of Sciences, Beijing 100049, China}
\affiliation{Songshan Lake Materials Laboratory, Dongguan, Guangdong 523808, China}

\author{Wu-Ming Liu}  \email{wliu@iphy.ac.cn}
\affiliation{Beijing National Laboratory for Condensed Matter Physics, Institute of Physics, Chinese Academy of Sciences, Beijing 100190, China}
\affiliation{University of Chinese Academy of Sciences, Beijing 100049, China}
\affiliation{Songshan Lake Materials Laboratory, Dongguan, Guangdong 523808, China}

\date{\today}
\begin{abstract}
We derive Lorentz-invariant four-fermion interactions, including Nambu-Jona-Lasinio type  and superconducting type, which are widely studied in high-energy physics, from the honeycomb lattice Hamiltonian with Hubbard interaction.
We investigate the phase transitions induced by these two interactions and consider the effects of the chemical potential and magnetic flux (Haldane mass term) on these phase transitions.
We find that the charge-density-wave and superconductivity generated by the attractive interactions are mainly controlled by the chemical potential, while the magnetic flux delimits the domain of phase transition.
Our analysis underscores the influence of the  initial topological state  on the phase transitions, a facet largely overlooked in prior studies. We present experimental protocols using cold atoms to verify our theoretical results. 
\end{abstract}
\maketitle
\section{Introduction}
Two-dimensional Dirac materials have emerged as a hot topic in the study of condensed matter physics. 
The electrons in these materials display a linear dispersion spectrum on the Fermi surface \cite{neto2009electronic}. 
Despite their fundamentally non-relativistic nature, their low-energy excitations can be described by the (2+1)-dimensional Dirac equation. 
This intriguing manifestation blurs the traditional boundaries between relativistic and non-relativistic realms in the microscopic world, indicating a possible transformation and connection through specific physical systems. However, the challenge lies in effectively bridging these critical aspects of condensed matter models with theoretical models in high-energy physics, thereby forging a pathway for relativistic studies in condensed matter physics. 
A particularly promising platform to explore this linkage is the Haldane-Hubbard model, setting the stage for the ensuing discussion in this work.

The Haldane model \cite{hhhaldane1988model}, a classical model of topological insulators, has attracted substantial attention \cite{neupert2011fractional,read2000paired,xu2011chern,yu2010quantized,thonhauser2006insulator}. This model describes a two-dimensional honeycomb lattice system with a non-zero Chern number, the characteristics of which have been experimentally verified over the past few years \cite{chang2013experimental}. 
The Haldane model with interaction has also been extensively studied \cite{herbut2006interactions,herbut2009theory,hickey2016haldane,vanhala2016topological,cangemi2019topological,yi2021interplay,miao2019exact,mai2023topological,yanes2022one,PhysRevLett.131.026601}. 
However, how to connect these essential, experimentally realizable condensed matter models with theoretical models in high-energy physics, and establish a bridge between condensed matter physics and high energy physics, remains an evolving field \cite{herbut2023so,herbut2023wilson,palumbo2014non,palumbo2013abelian,cirio20143}. 
In high-energy physics, four-fermion interactions play a crucial role in describing the properties of strongly interacting particles and understanding the low energy limit of the standard model. For example, the Nambu-Jona-Lasinio (NJL) model and Gross-Neveu (GN) model \cite{fernandez2021influence,fernandez2020renormalization,alves2017dynamical,khunjua2017inhomogeneous,khunjua2018dualities,khunjua2019charged,ebert1994effective,buballa2005njl,li2021gauge,ebert2001chromomagnetic,ebert2002influence,ebert2010cooper,ebert2011chiral,ebert2014competition,ebert2015interplay}, are used in the field of particle physics to study meson spectra, color superconductivity, and heavy ion collision physics, etc. 
And the  Thirring model \cite{khunjua2022hartree,gubaeva2022spontaneous,gies2010uv,gomes1991gauge}, is used to study dynamical symmetry breaking, and the phase diagram of real quantum chromodynamics, etc. Most of these models are phenomenological and lack experimental platforms. However, the unique properties of two-dimensional Dirac materials provide us with a unique opportunity to use these materials to simulate these models in high energy physics, particularly those involving interactions that exhibit Lorentz invariance.
In this paper, we commence with the Haldane model featuring Hubbard interactions and employ the van der Waerden notation to introduce a novel mapping from the honeycomb lattice Hamiltonian to a Lagrangian that encompasses two distinct interaction types: NJL-type and superconducting-type. This notation facilitates the transformation of the non-relativistic Hubbard interaction into relativistic four-fermion interactions, thus enabling the construction of a fully Lorentz-invariant Lagrangian. We subsequently investigate phase transitions induced by these two interactions by calculating the effective potential. Additionally, we explore the influence of chemical potential and the Haldane mass term (magnetic flux) inherent in the Haldane model on these phase transitions.
The Haldane mass term, as widely recognized, has gathered substantial attention in the realm of condensed matter physics and holds significance in high-energy physics. Scholarly works, such as those found in \cite{giuliani2016topological,yi2021interplay,vanhala2016topological,PhysRevLett.131.026601}, have delved deeply into the Haldane model in the context of repulsive Hubbard interactions. Simultaneously, the field of high-energy physics, including \cite{dudal2018remarks,olivares2020influence}, has explored the connections between the Haldane mass term and Chern-Simons interactions.
Our research centers on elucidating the interplay between the topological aspects of the Haldane model and phase transitions induced by weak Hubbard interactions. This area remains relatively unexplored in the existing literature \cite{core,klimenko2012superconducting,ebert2015interplay,gomes2021tilted,gomes2022tilted}, and our findings offer fresh insights. Notably, we have uncovered that the system, under attractive interactions, can exhibit superconducting and charge density wave (CDW) phases. Furthermore, certain regions of these phase transitions exhibit characteristics that are topologically protected. The pivotal role of the Haldane mass term in influencing the topological attributes of the model within this framework cannot be overstated.
Importantly, the Haldane mass term introduced in our study is not an adhoc parameter but naturally emerges from the effective theory of the Haldane model, originating from a physically meaningful quantity associated with magnetic flux:\(m_o = 3\sqrt{3}t'\sin\phi\).
This guarantees the theoretical consistency and practical applicability of our model, offering a fresh perspective on the direct relationship between phase transitions and topology in complex interacting systems.
Finally, in the Appendix \ref{experiments}, we propose experimental methods for probing these quantum phase transitions using cold atoms.

The rest of the paper is organized as follows. Section \ref{l2} introduces the mapping of the Haldane-Hubbard model to its corresponding Lagrangian, which incorporates both SC-type and NJL-type four-fermion interactions, utilizing van der Waerden notation. Section \ref{l3} is dedicated to the derivation of the system's effective potential. The renormalization  analysis of this effective potential is the focus of Section \ref{l4}, where we also present the resultant phase diagram. In Section \ref{l5}, we explore the role of the system's initial topological states in modulating interaction-induced phase transitions. Comprehensive derivations of equations cited in the main text can be found in the Appendix.

\section{Constructing the Lorentz-invariant Lagrangian} \label{l2}
\begin{figure}[h]
    \centering 
    \includegraphics[width=0.4\textwidth]{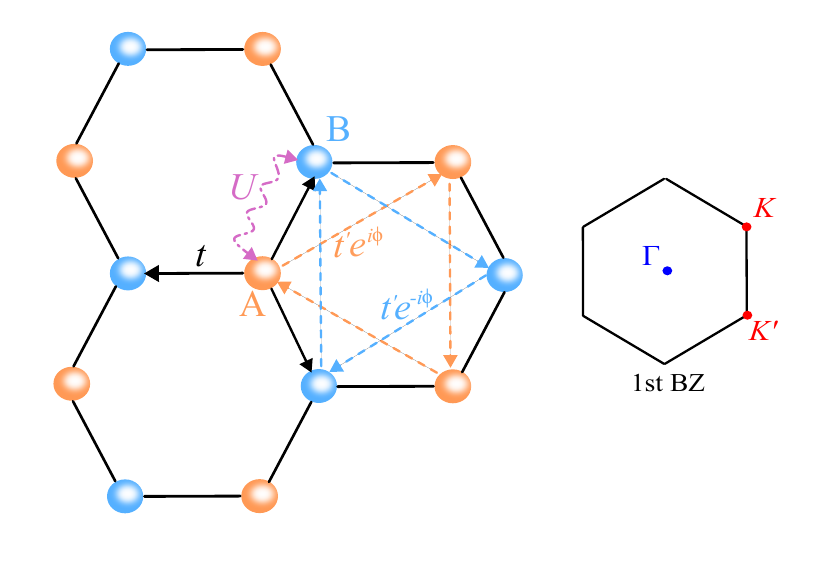} 
    \caption{ Haldane model on a honeycomb lattice with nearest neighbor interactions.
    The sublattices A and B are represented by orange and blue sites, respectively, each with an energy offset denoted by $\mu$. The real isotropic values are characterized by nearest-neighbor (NN) hopping terms $t$. The next-nearest-neighbor (NNN) hopping $t'$ is indicated by dashed lines with arrows, incorporating a phase factor $e^{\pm i\phi}$. The interaction $U$ is observed between adjacent particles. The diagram in the  right  delineates the Brillouin zone, featuring the Dirac points $K$ and $K'$.}
     \label{Fig.h} 
\end{figure}

We study a honeycomb lattice system with  nearest neighbor interactions, contributing to the following spinless Haldane-Hubbard model:
\begin{equation}
    \mathcal{H}=\mathcal{H}_0+\mathcal{H}_T+\mathcal{H}_S+\mathcal{H}_U  \,, \label{eq-hh}
\end{equation}
with partial Hamiltonian expressed as:
\begin{align}
  \mathcal{H}_0 &= -\sum_{\langle i, j \rangle} t (a_i^{\dagger} b_j+\text{H.c.}), \\
  \mathcal{H}_T &= -\sum_{\langle \langle i, j \rangle \rangle} t' (e^{i \phi} a_i^\dag a_j+e^{-i \phi} b_i^\dag b_j), \\
  \mathcal{H}_S &= \sum_{i\in A} \mu a_i^\dag a_i-\sum_{i\in B} \mu b_i^\dag b_i, \\
  \mathcal{H}_U &= - \sum_{\langle i j\rangle} U a^\dag_i a_i b^\dag_j b_j.
  \end{align} 
Here the summation $\sum_{\langle i,j \rangle}$ takes over all nearest-neighbor (NN) sites,
and $\sum_{\langle \langle i, j \rangle \rangle}$ takes over all next-nearest-neighbor (NNN) sites.
$t$ and $t'$ are real-valued NN and NNN hopping amplitudes,
and the latter contain addtional phase $\pm \phi$ for different sublattices along the arrows shown in Fig. \ref{Fig.h}.
The particle creation and annihilation operators are denoted by $a^\dag_i(b^\dag_i)$ and $a_i(b_i)$ for A(B) sublattice.
The energy offset $\mu$ between sites of A-B sublattices breaks inversion symmetry.
$U$ denotes the NN interaction strength between particles in different sublattices.
In the following we explain how the Haldane model can be associated with free dirac fermion Lagrangian, and
gives an exact mapping from the Hubbard interaction to those Lorentz invariant four-fermion interactions.

We focus on the NN hopping Hamiltonian $\mathcal{H}_0$. 
From the dispersion relation for the noninteracting theory we obtain six $K$ points, and our choice of two inequivalent ones, which we denote $K_{\pm}$, these correspond to the $K$ and $K'$ points in Fig.~\ref{Fig.h}.
Expanding around $K_{\pm}$, the Hamiltonian $\mathcal{H}_0$ becomes: $H_0\!=\!\int\!\frac{d^2 p}{(2\pi)^2}(\psi_{+}^{\dagger}(p) v_{F} \sigma \cdot p \psi_{+}(p)\!-\!v_{F} \psi_{-}^{\dagger}(p)\left(p_{x} \sigma_{x}-p_{y} \sigma_{y}\right) \psi_{-}(p))$, where
$\psi_{\pm}(p)\!=\!(a(K_{\pm}+p),b(K_{\pm}+p))^\intercal$, $v_F=3td/2$ ($d$ is the lattice constant). And afterwards, for convenience, we will denote $a(K_{\pm}+p)$ by $a_{\pm}$, and $b(K_{\pm}+p)$ by $b_{\pm}$.

We encapsulate the Hamiltonian of point $K$ and point $K^{\prime}$ inside a four-component spinor $\Psi^{\intercal}=\left(a_{+} , b_{+} , b_{-} , a_{-}\right)$.
After using the Legendre transformation we can obtain the qiLagrangian of the system: $\mathcal{L}=\bar{\Psi}\left(i \gamma^{v} \partial_{v}\right) \Psi$.
where $\bar{\Psi}=\Psi^{\dagger}\gamma^{0}$, $v=0,1,2$ and we set $v_F=1$ for convenience. We use the irreducible four-dimensional spinor representation, the gamma matrices are $\gamma^{\mu}= \left(\begin{array}{cc}
    0 & \bar{\sigma}^\mu \\
    \sigma^{\mu} & 0
    \end{array}\right)$
and $\mu=0,1,2,3$. There exist one other matrice $\gamma^{5}= \left(\begin{array}{cc}
    1 & 0 \\
    0 & -1
    \end{array}\right)$, which anticommute with all $\gamma^{\mu}$ (see Appendix \ref{aa}).
 We now consider the NNN hopping $\mathcal{H}_T$ and the chemical potential term $\mathcal{H}_S$.
 The effective Lagrangian obtained by expanding $\mathcal{H}_T+\mathcal{H}_S$  at the Dirac point $K,K^{\prime}$ are $\mathcal{L}_m=\psi_{\pm}^{\dagger} m_{\pm} \sigma_{z} \psi_{\pm}$.
Where $m_{\pm}=\mu \pm 3 \sqrt{3} t^{\prime} \sin \phi$. For convenience, we set $m_o=3 \sqrt{3} t^{\prime} \sin \phi$, and it is actually the Haldane mass term \cite{carrington2019effect,dudal2018remarks,gusynin2007ac}.
 Now, encapsulate it inside the four-component spinor, we can get the on-site  and Haldane mass terms respectively:  $\mu \bar{\Psi}\gamma^3 \Psi$, $m_o\bar{\Psi} \gamma^3 \gamma^5 \Psi$ (see Appendix \ref{cc}).
 Finally, we can get the free Dirac fermion Lagrangian from the effective theory of Haldane model is: $\mathcal{L}=\bar{\Psi}\left(i \gamma^{v} \partial_{v} +\mu \gamma^3 + m_o \gamma^3 \gamma^5 \right) \Psi$ .

We now consider the Hubbard interaction $H_{U}$ in continue limit and ignore the interaction between $K$ and $K^{\prime}$ \cite   {luo2015collective,luo2015comment}.
We take into account the Hubbard interaction for both $K$ and $K^{\prime}$ point, i.e.,  $H_{U} \approx H_{int}\equiv \int d {\bf r} \ U (a_{+} ^{\dagger}a_{+} b_{+} ^{\dagger}b_{+}+a_{-} ^{\dagger}a_{-}b_{-} ^{\dagger}b_{-}) \label{8}$ .
Following, we'll use the  van der Waerden notation to map the $H_{int}$ to the Lorentz invariant four-fermion interactions, which contain NJL-type and Superconducting-type.
Below we will demonstrate how to create this mapping.
We note that $\Psi=\left(\begin{array}{l}\psi_{+} \\ \sigma_1\psi_{-}\end{array}\right)$, and for convenience we define $\chi=\sigma_1 \psi_{-}$ and $\eta=\psi_{+}$. Actually, $\eta$ is the right-chiral weyl spinor, and the corresponding $\chi$ is the left-chiral spinor. 
Thus, $\Psi$  is indeed a Dirac spinor containing two different chiral weyl spinors, i.e., $\Psi=\left(\begin{array}{l}\eta \\ \chi\end{array}\right)$. 
We introduce the van der Waerden notation and since we are using the path integral framework, the elements inside $\Psi$ are grassman numbers rather than operators. According to this notation \cite{muller2010introduction,labelle2010supersymmetry,martin2010supersymmetry},  we have 
$ \chi_1=b_-, \; \chi_2=a_-,\;\chi^a=\epsilon^{ab}\chi_b ,\;\bar{\eta}^{\dot{1}}=a_+  , \;\bar{\eta}^{\dot{2}}=b_+ ,\;\bar{\eta}^{\dot{a}}=\epsilon^{\dot{a}\dot{b}}\bar{\eta}_{\dot{b}}$,
where $\epsilon^{12}=\epsilon^{\dot{1}\dot{2}}=1$ and $\epsilon_{12}=\epsilon_{\dot{1}\dot{2}}=-1$. We use the notation $\chi \cdot \chi \equiv \chi^a \chi_a=\chi^{\intercal}(-i\sigma_2 )\chi$ and $\bar{\chi} \cdot \bar{\chi}\equiv \bar{\chi}_{\dot{a}} \bar{\chi}^{\dot{a}}=\chi^{\dagger}i\sigma_2\chi^{\dagger \intercal}$, which are invariant under Lorentz transformation.
The Hamiltonian density of the Hubbard interaction can be expressed as: $4 a_{+} ^{\dagger}  a_{+} b_{+} ^{\dagger} b_{+}\!=\!( \eta \cdot \eta)(\bar{\eta} \cdot \bar{\eta})$.
Similarly, $4 a_{-} ^{\dagger} a_{-} b_{-} ^{\dagger}  b_{-}\!=\!(\bar{\chi} \cdot \bar{\chi})(\chi \cdot \chi)$. We then introduce a charge conjugate operator $C=-i\gamma^2 \gamma^0$ which satisfies 
$C=-C^{-1}=-C^{\dagger}$. So, the Hubbard interaction  finally can be mapped to the Lorentz invariant four-fermion interactions: $H_{int}\!=\!\frac{U}{4}[ (\Psi^{\intercal} C \Psi)(\bar{\Psi} C \bar{\Psi}^{\intercal})\!+\!(\bar{\Psi}\Psi)^2\!-\!(\bar{\Psi}i\gamma^5\Psi)^2]$ (see Appendix \ref{cc}).

In the following, we discuss the properties of the model in a more general sense, i.e., using the following Lagrangian:
\begin{align}
    \mathcal{L}&=\sum_i \bar{\Psi}_i \left(i \gamma^{v} \partial_{v} +\mu \gamma^3 + m_o \gamma^3 \gamma^5 \right) \Psi_i \notag   \\ 
    &+\frac{G_1}{N}[(\sum_i\bar{\Psi}_i\Psi_i)^2-(\sum_i \bar{\Psi}_i i \gamma^5\Psi_i)^2] \notag   \\
    &+\frac{G_2}{N}(\sum_i  \Psi^{\intercal}_i C \Psi_i) \sum_j(\bar{\Psi}_j C \bar{\Psi}^{\intercal}_j)   \,.
\end{align}
Here, we have adopted the large-N assumption that all fermion fields $\Psi_i (i = 1,...,N)$ form a fundamental multiplet
of the $O(N)$ group. 
This model is similar to those investigated in Refs. \cite{core,klimenko2012superconducting,ebert2015interplay,gomes2021tilted,gomes2022tilted}, but we consider the effect of the haldane mass term, an aspect not addressed in these references. Our assumption $G_1=G_2=\frac{U}{4}$ are derived from Hubbard interaction rather than a direct phenomenological parameter, and all the order parameters representing phases that we discuss subsequently originate from genuine condensed matter systems. 
It is worth mentioning that in this paper, we adopt a distinct gamma matrix representation. 
However, as we will demonstrate below, the physical results are independent of the representation chosen for the gamma matrices.
And in the main text, we will only consider the parts that are directly related to the Hubbard interaction, i.e., $G_1=G_2=G$.

\section{Effective potential} \label{l3}
We use the Hubbard-Stratonovich (HS) transformation to decouple these four-fermion interactions and solve for the thermodynamic potential (TDP). 
We focus on the case with Hubbard interaction, where we set $G_1=G_2=G$ and $G=U/4$. 
Introducing the auxiliary fields $\Delta,\pi,\sigma$, we have:
\begin{align}
&\mathcal{L}=\bar{\Psi}_i \left(i \gamma^{v} \partial_{v} +\mu \gamma^3 + m_o \gamma^3 \gamma^5 -\sigma-i\gamma^5 \pi \right) \Psi_i \notag \\
&-\frac{N}{4G}(\sigma^2+\pi^2)-\frac{N}{4G}\Delta^{\ast}\Delta-\frac{\Delta^{\ast}}{2}\Psi^{\intercal}_i C \Psi_i-\frac{\Delta}{2}\bar{\Psi}_i C \bar{\Psi}^{\intercal}_i.
\end{align}
Using the Euler-Lagrange equations of motion for these auxiliary fields which take the form:
$\sigma(x)=-2 \frac{G}{N} (\bar{\Psi}_i \Psi_i)$, $\pi(x)=-2 \frac{G}{N} (\bar{\Psi}_i i \gamma^5 \Psi_i)$ and  $\Delta(x)=-2 \frac{G}{N} (\Psi_i ^\intercal C \Psi_i)$.
The ground state expectation values $\langle \Delta(x) \rangle,\langle \pi(x) \rangle,\langle \sigma(x) \rangle$ of the composite bosonic fields are determined by
the saddle point equations:
\begin{equation}
  \frac{\delta {\cal S}_{\rm {eff}}}{\delta\sigma (x)}=0,\;\frac{\delta {\cal S}_{\rm {eff}}}{\delta\Delta (x)}=0,\; \frac{\delta {\cal S}_{\rm {eff}}}{\delta\pi (x)}=0.
\end{equation}
Where
  \begin{equation} 
    \exp(i {\cal S}_{\rm {eff}}(\sigma,\Delta,\Delta^{*}))=
\int\prod_{l=1}^{N}[d\bar\psi_l][d\psi_l]\exp\Bigl(i\int {\cal
L}\,d^3 x\Bigr).
  \end{equation}
For simplicity, throughout the paper we suppose that the above mentioned ground state expectation values do not depend on space-time coordinates: $ \vev{\sigma(x)}  \equiv \sigma$, $\vev{\pi(x)} \equiv \pi$, $    \vev{\Delta(x)}\equiv \Delta$.
So, in the leading order of the large-$N$ expansion, after integrating out the fermionic field, we can derive the TDP:
\begin{equation} \label{19}
\Omega=\frac{\sigma^2+\pi^2}{4G}+\frac{\Delta^2}{4G}+\frac{i}{2}\sum_i \int \frac{d^3 p}{(2\pi)^3}log \lambda_i (p) \,,
\end{equation}
where $\Delta$ is a real number, $\lambda_i (p)$($i=1,2,3,4$) is the four roots of the four-by-four matrix $\mathcal{D}(p)=-\Delta^2 \bf{I}+\mathcal{D}_{+} \mathcal{D}_{-}$ (see Appendix \ref{bb}).
Without loss of generality, for chiral symmetry breaking, we can always choose a direction such that $\pi=0, \sigma \neq 0$.
In powers of $p_0$, we can write $\prod_i \lambda_i (p)=P^+(p_0)P^- (p_0)$. According to the general theorem of algebra,
the polynomial $P^{\beta} (p_0)$($\beta=+,-$) can be presented in the form: 
  \begin{equation} 
    P^{\beta} (p_0)=(p_0-p^{\beta} _{01})(p_0-p^{\beta} _{02})(p_0-p^{\beta} _{03})(p_0-p^{\beta} _{04}).
  \end{equation}
Then,  the TDP is: 
  \begin{equation} 
    \Omega=\frac{\sigma^2}{4G}+\frac{\Delta^2}{4G}+\frac{i}{2} \sum_{\beta} \int \frac{d^3 p}{(2\pi)^3}log P^{\beta} (p_0).
  \end{equation}
Using  the  general formula: $\int_{-\infty}^\infty dp_0\ln\big(p_0-R)=\mathrm{i}\pi|R|$.
Where R is a real quantity, it is possible to reduce the TDP to the following: $\Omega \equiv \Omega^{un}(\Delta,\sigma)=\frac{\sigma^2}{4G}+\frac{\Delta^2}{4G} -\frac{1}{4} \sum_{\beta,i} \int \frac{d^2 \bm{p}}{(2\pi)^2}(|p^{\beta} _{0i}|)$.

\begin{figure*}[t]
    \centering 
    \includegraphics[width=0.98\textwidth]{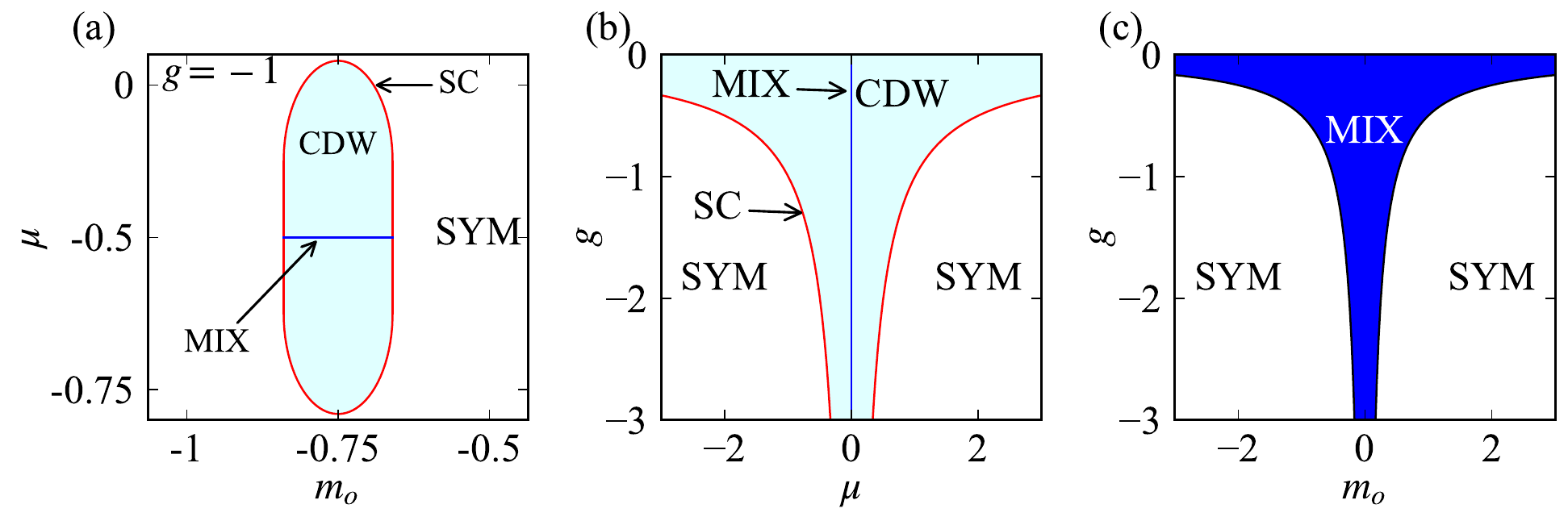} 
    \caption{The phase diagrams of the system under various fixed parameters: (a) The phase diagram in the $(\mu, m_o)$ parameter space, where the interaction strength $g = -1$.
    (b) The phase diagram in the $(g, \mu)$ parameter space, where the Haldane mass term $m_o=0$. (c) The phase diagram in the $(g, m_o)$ parameter space, where the chemical potential $\mu = 0$.
    In the diagrams, the white area denotes the system in a symmetry (SYM) phase, the sky-blue area represents the system in a CDW phase, the red area indicates the system is in a superconductor (SC) phase, and the blue area signifies the system is in a mixed (MIX) phase of CDW and superconductivity.} 
\label{f1s}
\end{figure*}

\section{Renormalization and phase diagram}  \label{l4}
First, we will obtain a finite (i.e., renormalized) expression for the TDP at $m_o=0$ and $\mu=0$. That is, in vacuum: 
  \begin{align} 
    &V^{un}(\Delta,\sigma)=\frac{\sigma^2}{4G}+\frac{\Delta^2}{4G}  \notag \\
    &-\int \frac{d^2 \bm{p}}{(2\pi)^2} (\sqrt{\bm{p}^2+(\Delta+\sigma)^2}+\sqrt{\bm{p}^2+(\Delta-\sigma)^2}).
  \end{align}
Next, let us regularize the effective potential by cutting momenta; we suppose that $|\bm{p}|<\Lambda$. (Note here that we take a spherical coordinate truncation instead of the square truncation in Ref. \cite{core}, so there is a difference in the sense of constants). As a result, we have the following regularized expression, which is finite at finite values of $\Lambda$:
\begin{align} \label{30}
V^{un}(\Delta,\sigma)&=(\sigma^2+\Delta^2) (\frac{1}{4G}-\frac{\Lambda}{2\pi})  -\frac{\Lambda^3}{3\pi} \notag \\ 
&+\frac{1}{6\pi}(|\Delta+\sigma|^3+|\Delta-\sigma|^3)  \,.
\end{align}
We assume that the bare
coupling constants $G$  depend on the cutoff parameter
of $\Lambda$ in the way that in the limit
$\Lambda \to \infty$ yields a finite expression in square brackets (\ref{30}) . Obviously, in order to satisfy this requirement,
it is sufficient to require that $\frac{1}{4G} \equiv \frac{1}{4G (\Lambda)}=\frac{\Lambda}{2\pi}+\frac{1}{2\pi g}$.
where $g$ is finite and $\Lambda$-independent model parameter with dimensionality of inverse mass.
Ignore there an infinite $\sigma$- and $\Delta$-independent constant $-\frac{\Lambda^3}{3\pi}$,
one obtains the following renormalization, i.e., finite expression
for the effective potential: $ V^{ren}(\Delta,\sigma)=\lim_{\Lambda\to\infty}
\left\{V^{un}(\Delta,\sigma)\Big |_{G(\Lambda)}+\frac{\Lambda^3}{3\pi}\right\}$.
It should also be mentioned that the $V^{ren}$ is a
renormalization group invariant quantity. Finally, we get:
  \begin{equation}
    V^{ren}(\Delta,\sigma)=\frac{\sigma^2}{2 \pi g}+\frac{\Delta^2}{2 \pi g} 
+\frac{1}{6\pi}(|\Delta+\sigma|^3+|\Delta-\sigma|^3).
  \end{equation}
The coordinates of the global minimum point
$(\Delta_0,\sigma_0)$ of the effective potential $V^{ren}(\Delta,\sigma)$ define the ground
state expectation values of auxiliary fields $\sigma (x)$ and
$\Delta (x)$. Namely, $\sigma_0=\vev{\sigma(x)}$ and
$\Delta_0=\vev{\Delta(x)}$. The quantities $\sigma_0$ and $\Delta_0$ are
usually called order parameters, or gaps. Moreover, the gap $\sigma_0$ is equal to the dynamical
mass of  one-fermionic excitations of the ground state. That is, its appearance is associated with chiral symmetry breaking.
In this work, originating from genuine condensed matter systems, the specific pairing of $\sigma$ arises from the particle-hole pairing located at two inequivalent Dirac points, leading to the charge-density-wave (CDW) (see Appendix \ref{cc}).
The emergence of $\Delta_0$ is associated with the onset of the superconducting phase transition.
It is worth noting that although the regularization scheme we adopt and the representation of the gamma matrix are different from those in Ref.~\cite{core}, the final resulting effective potential is indeed precisely the same, so does the phase portrait.
This demonstrates that the effective potential is indeed a renormalization group invariant quantity, and it confirms that the physics of the system is independent of the representation chosen for the gamma matrices.

We now study the influence of the Haldane mass term $m_o$ and the chemical potential $\mu$ on the phase structure of the model.
We have (see Appendix \ref{bb})
\begin{align}  \label{fullpotential}
    &\Omega^{ren}(\Delta,\sigma) = \frac{\sigma^2}{2 \pi g} + \frac{\Delta^2}{2 \pi g} +\frac{1}{12\pi} (|\Delta+m_o+\sqrt{\mu^2+\sigma^2}|^3\notag \\
    &+|\Delta+m_o-\sqrt{\mu^2+\sigma^2}|^3+|\Delta-m_o+\sqrt{\mu^2+\sigma^2}|^3  \notag \\
    &+ |\Delta-m_o-\sqrt{\mu^2+\sigma^2}|^3) .  
\end{align}

\begin{figure*}[t]
    \centering 
    \includegraphics[width=0.98\textwidth]{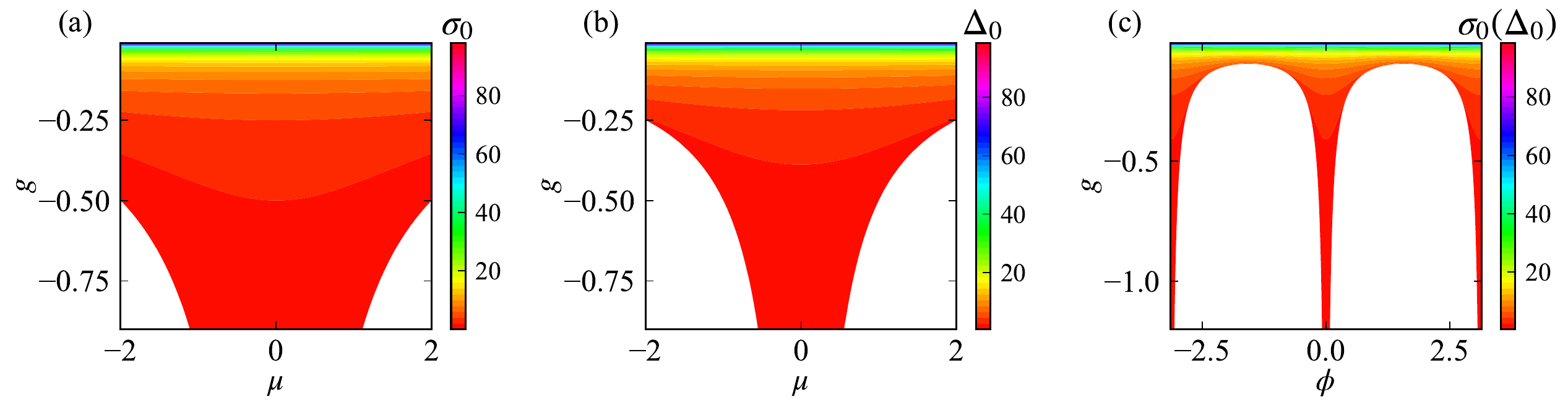} 
    \caption{Thermal diagrams of order parameters versus interaction $g$ and chemical potential $\mu$ or magnetic flux $\phi$ (Haldane mass $m_o$), respectively.
    (a,b) Thermal diagrams of the order parameters $\sigma_0$, $\Delta_0$ in the parameter space ($g, \mu$), respectively. (c) Thermal diagrams of the order parameters $\sigma_0$, $\Delta_0$ in the parameter space ($g, \phi$), where $\sigma_0 = \Delta_0$.  }
\label{f2s}
\end{figure*}

By solving the extreme points of the given TDP, we can derive the system's phase diagram, which is determined by the interplay of three parameters ($g, \mu, m_o$). 
We find that in the repulsive region ($g>0$), no spontaneous symmetry breaking occurs. 
Conversely, in the attractive region ($g<0$), the system undergoes spontaneous symmetry breaking, leading to a transition from the CDW to superconductivity.
Similar to previous studies that have incorporated chemical potential effects in the context of attractive Hubbard interactions \cite{zhao2006bcs,iskin2019superfluid,cichy2018reentrant}, the chemical potential in our model also induces superconducting phase transitions. However, the region exhibiting superconductivity in our model is quite peculiar, manifesting only at the phase boundary. Furthermore, the chemical potential 
$\mu$ plays a pivotal role in inducing transitions to both the superconducting phase and the charge density wave (CDW).
Meanwhile, the Haldane mass term merely triggers the system's transition process from a symmetric phase to a mixed phase.
The corresponding phase diagrams are shown in Fig. \ref{f1s}.

Next, we plotted the phase diagrams of the $(g,m_o)$ and $(g,\mu)$ systems, where $(\mu=0, m_o=0)$ correspond to Fig. \ref{f1s} (b and c), and the equations of the corresponding boundary curves of these two plots are $g=-\frac{1}{|\mu|}$ and $g=-\frac{1}{2|m_o|}$, respectively. We find that when $\mu=0$, under attractive interaction, tuning $m_o$ from negative to positive leads the system to undergo phase transitions from symmetric phase to mixed phase and back to symmetric phase. On the other hand, at $m_o=0$, adjusting the chemical potential $\mu$ induces the system to experience phase transitions from a symmetry phase to a superconducting phase, followed by the emergence of CDW (apart from the line at $\mu = 0$, where the system is in a mixed phase) and then back to symmetry. This suggests that the chemical potential $\mu$ strongly induces the emergence of CDW in the system, while the Haldane mass term merely restricts the range of symmetry breaking.
In the Appendix \ref{appen4}, we also draw the phase diagrams for ($g,\mu$) and  ($g,m_o$) with $m_o$ and $\mu$ not equal to 0, respectively.
In order to compare with the experiment, we also plotted the order parameters as a function of magnetic flux and chemical potential, see Fig. \ref{f2s}, where (a) and (b) is the thermal diagrams of $\sigma_0(g,\mu)=\sqrt{1/g^2 - \mu^2}$ and $\Delta_0(g,\mu) =\frac{-1 / g + \sqrt{1 / g^2 - 4\mu^2}}{2}$, respectively ($m_o=0$ in this case). (c) is the thermal diagrams of $\sigma_0(g,\phi)=\Delta_0(g,\phi)=\frac{-1 / g + \sqrt{1 / g^2 - 4 (3\sqrt{3}t\sin\phi)^2}}{2}$ (where we have set $t=1$ and $\mu=0$).


\section{Unveiling the Correlation: Initial Topological States and $(g_1, g_2)$ Interaction-Induced Phase Transitions} \label{l5}
In this Section, we undertake a comprehensive analysis of the system's phase diagram in the $(g_1, g_2)$ parameter space,
we analyze the phase structure in the planes defined by the parameters ($g_1,g_2$). Furthermore, we will investigate the relationship between the initial topological state of the system and the phase transitions induced by interactions. Remarkably, this specific aspect remains unexplored in the existing literature.
We recognize that the topology of the Haldane model is determined by the Chern number: $C=\frac{1}{2} (\text{sgn}(\mu+m_o)-\text{sgn}(\mu-m_o))$, which depends on the relative magnitude of the chemical potential and the Haldane mass term. 
Specifically, when $\mu>m_o$, $C=0$ and the system resides in a topologically trivial state; when $\mu<m_o$, $C=1$ and the system becomes a topological insulator (Due to the symmetry of the thermodynamic potential (\ref{fullpotential}), i.e., it remains invariant when \( \mu \rightarrow -\mu \) and \( m_o \rightarrow -m_o \), we limit our discussion to the cases with \( m_o > 0 \), \( \mu > 0 \), and \( C \geq 0 \)). We first discuss two extreme scenarios: 1. $m_o\neq0$, $\mu=0$; 2. $m_o=0$, $\mu\neq0$. These cases correspond to two distinct topological states ($C=1$, $0$). 
Using Eq. (\ref{fullpotential}), we derive the respective thermodynamic potentials (TDP) as follows:
\begin{align}
  V_{m_o} &\equiv \Omega^{ren}(\Delta,\sigma,m_o) \notag \\
  &= \frac{\sigma^2}{2 \pi g_1} + \frac{\Delta^2}{2 \pi g_2}+\frac{1}{12\pi} (|\Delta+m_o+\sigma|^3 \notag \\
             &+ |\Delta+m_o-\sigma|^3 +|\Delta-m_o+\sigma|^3 + |\Delta-m_o-\sigma|^3),\notag \\
  V_{\mu} &\equiv \Omega^{ren}(\Delta,\sigma,\mu)\notag \\
  & = \frac{\sigma^2}{2 \pi g_1} + \frac{\Delta^2}{2 \pi g_2}  \notag \\
        &+\frac{1}{6\pi} (|\Delta+\sqrt{\mu^2+\sigma^2}|^3 + |\Delta-\sqrt{\mu^2+\sigma^2}|^3)   \,.
\end{align}
Solving the system of equations,    $\begin{cases}
    \frac{\partial V_i}{\partial \Delta} = 0,\\
    \frac{\partial V_i}{\partial \sigma} = 0.  
    \end{cases}$($i=m_o,\mu$),
we analytically determine the system's order parameters and the conditions for phase boundaries in these two extreme scenarios. The corresponding phase diagrams are depicted below.
\begin{figure}[h]
  \centering
  \includegraphics[width=0.5\textwidth]{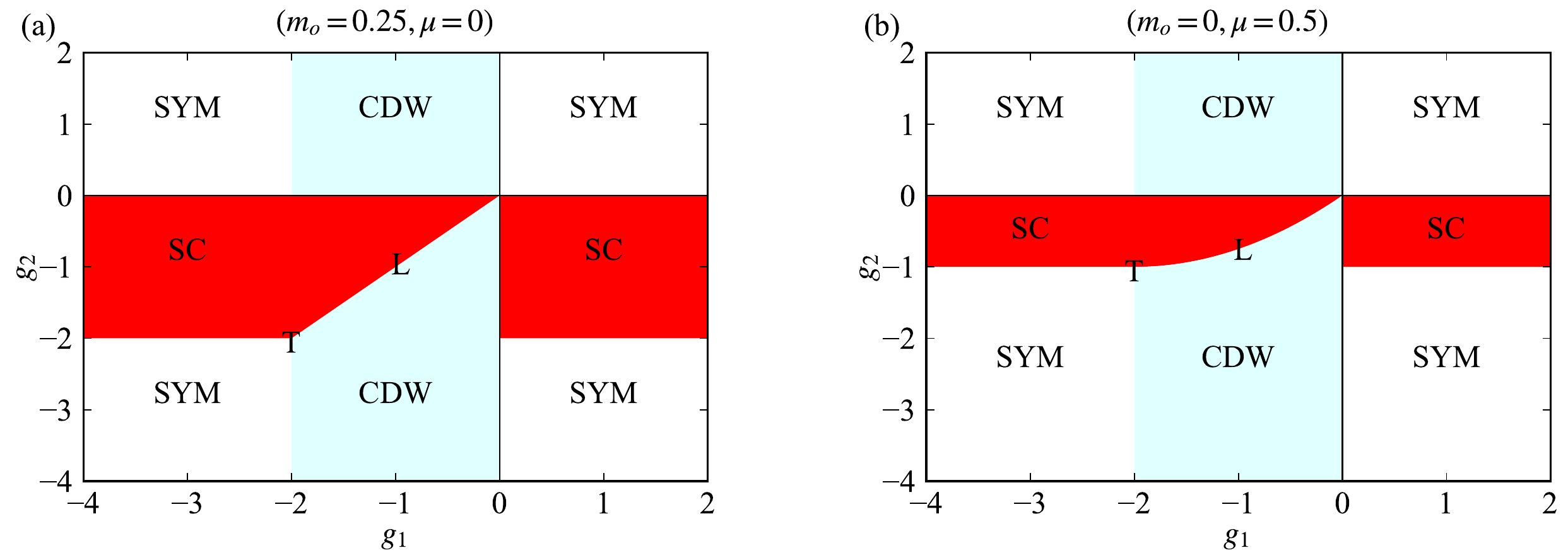}
   \caption{The ($g_1$,$g_2$)-phase portrait of the model. 
  The T point denotes the tricritical point, and the line labeled as L represents the phase boundary in the third quadrant. (a): The ($g_1$,$g_2$)-phase portrait at ($m_o=0.25,\mu=0$). (b):  The ($g_1$,$g_2$)-phase portrait  at ($m_o=0$ and $\mu=0.5$). }
   \label{fig-S2}
\end{figure}

Fig. \ref{fig-S2} (a) presents the ($g_1$, $g_2$) phase diagram in the topological insulator state, where the analytically derived tricritical point T is located at $g_1=g_2=-\frac{1}{2|m_o|}$. The Superconducting order parameter $\Delta_0(m_o)=\frac{-\frac{1}{g_2}+\sqrt{\frac{1}{g_2 ^2}-4m_o ^2}}{2}$ and the CDW order parameter $\sigma_0(m_o)=\frac{-\frac{1}{g_1}+\sqrt{\frac{1}{g_1 ^2}-4m_o ^2}}{2}$. 
Fig. \ref{fig-S2} (b) depicts the ($g_1$, $g_2$) phase diagram in the topologically trivial state, with the tricritical point T located at $g_1=-\frac{1}{|\mu|}, g_2=-\frac{1}{2|\mu|}$. The Superconducting order parameter $\Delta_0(\mu)=\frac{-\frac{1}{g_2}+\sqrt{\frac{1}{g_2 ^2}-4\mu ^2}}{2}$ and the CDW order parameter $\sigma_0(\mu)=\sqrt{\frac{1}{g_1 ^2}-\mu ^2}$.

\begin{figure}[htbp]
  \centering
  \includegraphics[width=0.5\textwidth]{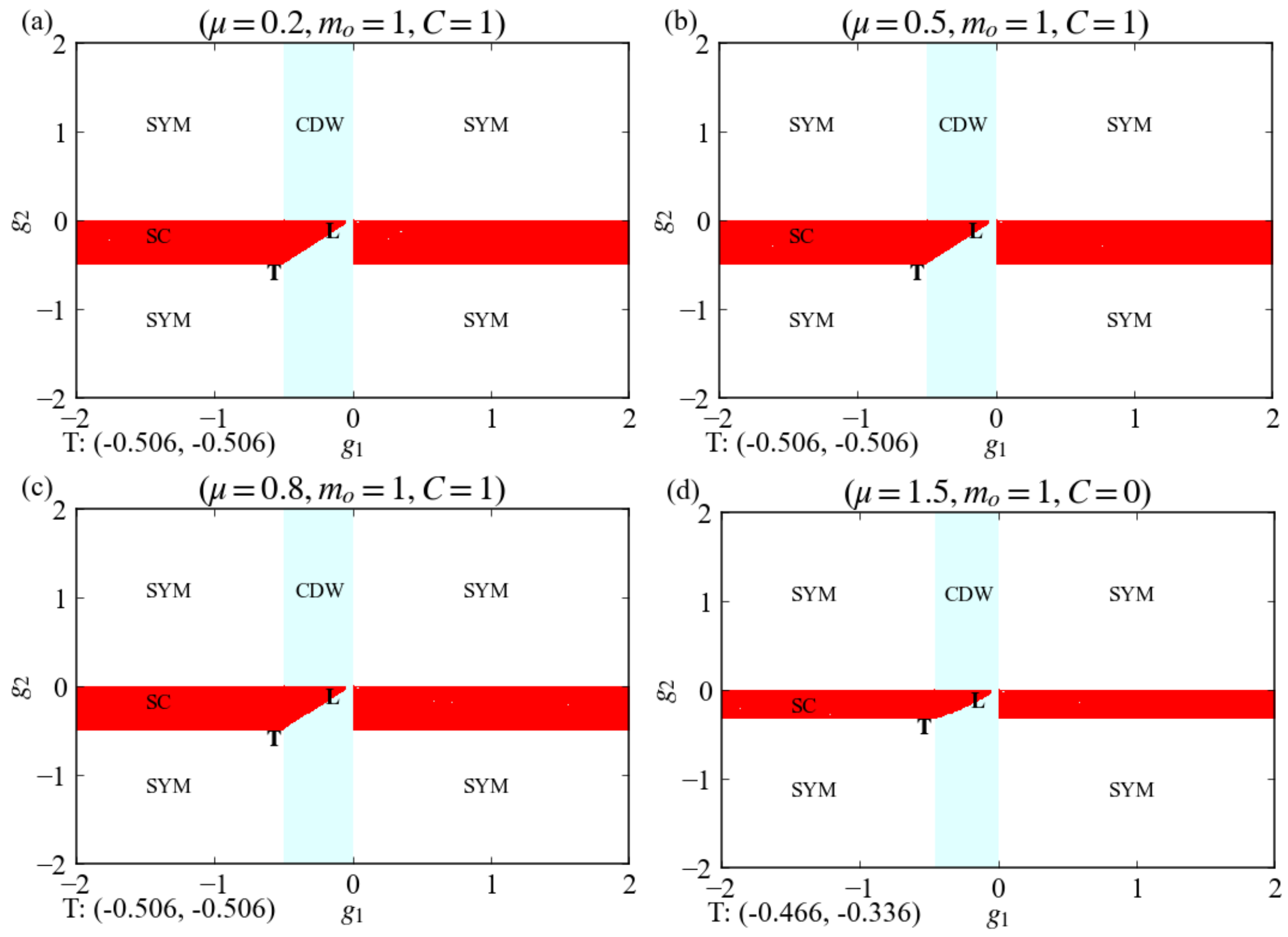}
  \caption{The  ($g_1,g_2$)-phase portrait  for fixed chemical potential and Haldane mass term. 
  Figures (a), (b), and (c) depict the phase diagrams for the topological insulator state $(C=1)$,
  while Figure (d) illustrates the phase diagram for the topologically trivial state. Each diagram is generated by holding the magnitude of $m_o$ constant and progressively increasing the magnitude of $\mu$. 
  The coordinates of the tricritical point T, computed numerically, are provided in the lower left corner of each figure.}
  \label{fig-S3}
\end{figure}

\begin{figure}[htbp]
  \centering
  \includegraphics[width=0.5\textwidth]{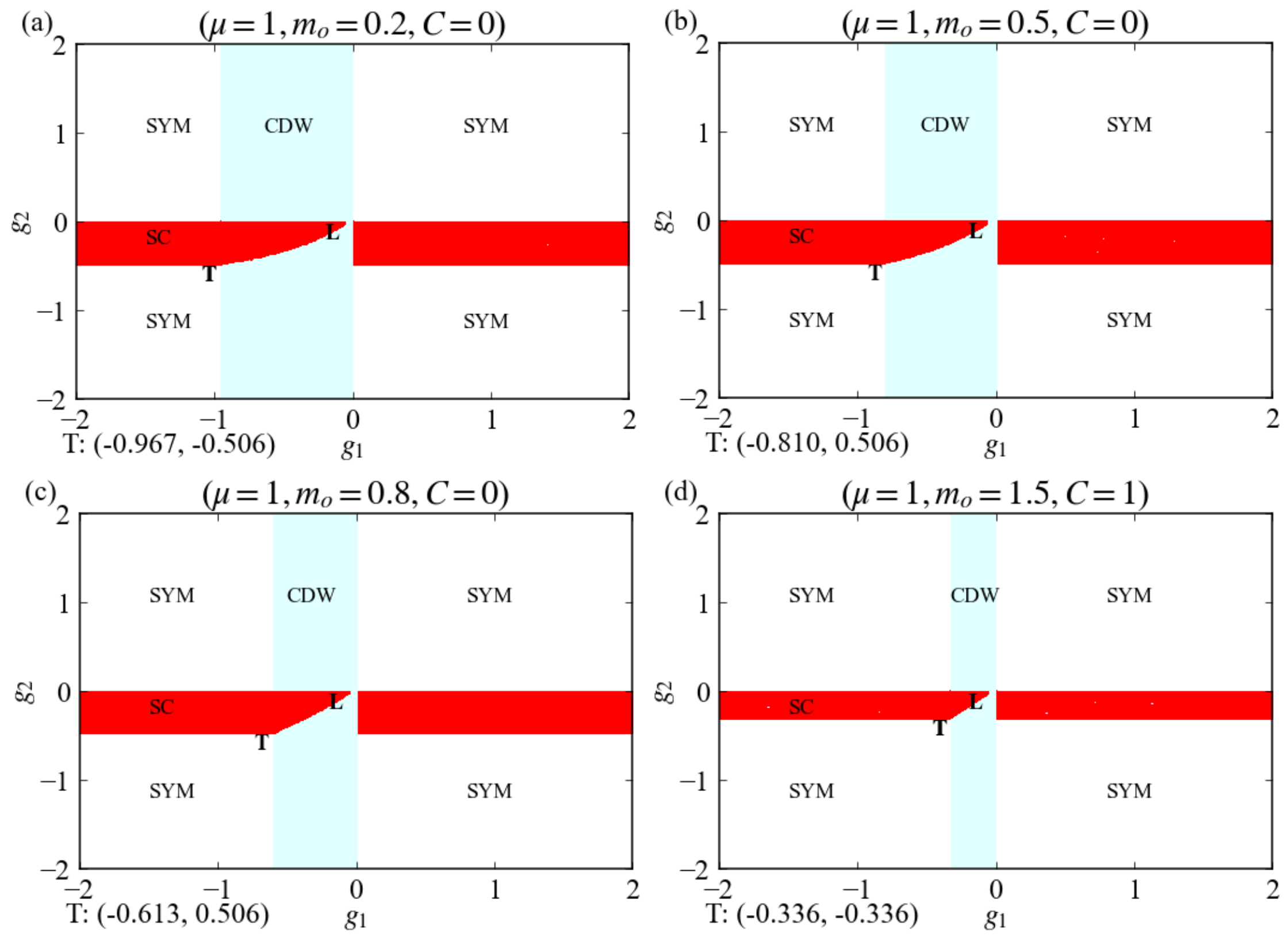}
  \caption{The  ($g_1,g_2$)-phase portrait for fixed chemical potential and Haldane mass term.
  Figures (a), (b), and (c) depict the phase diagrams for the topological trivial state $(C=0)$,
  while Figure (d) illustrates the phase diagram for the topologically insulator state. Each diagram is generated by holding the magnitude of $\mu$ constant and progressively increasing the magnitude of $m_o$.}
  \label{fig-S4}
\end{figure}

Upon comparing ($a$) and ($b$), it is evident that the expressions for the triple point T differ in these two extreme scenarios. This observation leads us to pose a question: Is the coordinate of the triple point T related to the topology of the system? Specifically, when the initial system is in a topological state, is the coordinate of T topologically protected?
The answer is affirmative; the coordinate of the triple point T is indeed topologically protected. We have demonstrated this by numerically solving the phase diagram of the system when both $\mu$ and $m_o$ are simultaneously non-zero, thereby proving that the coordinate of the triple point T is topology-dependent. The numerical phase diagram is presented in Fig. \ref{fig-S3} and \ref{fig-S4}.

When the system initially resides in a topological insulator state, i.e., $\mu<m_o$, the expression for the coordinates of the triple point in the $(g_1,g_2)$ plane remains constant at ($-\frac{1}{2|m_o|}$, $-\frac{1}{2|m_o|}$), irrespective of the value of the chemical potential $\mu$. Furthermore, the phase boundary L is a straight line.

Conversely, when the system initially resides in a topologically trivial state, i.e., $\mu>m_o$, the expression for the coordinates of the triple point in the $(g_1,g_2)$ plane becomes ($-\sqrt{\frac{1}{\mu^2+2m_o ^2}}$,$-\frac{1}{2|\mu|}$), and the phase boundary L becomes a curve.

Therefore, the topology of the system not only protects the coordinates of the triple point T induced by interaction-induced phase transitions but also preserves the geometric shape of the phase boundary L. Our conclusions are summarized in the subsequent table.
\begin{figure}[H]
  \centering
  \includegraphics[width=0.5\textwidth]{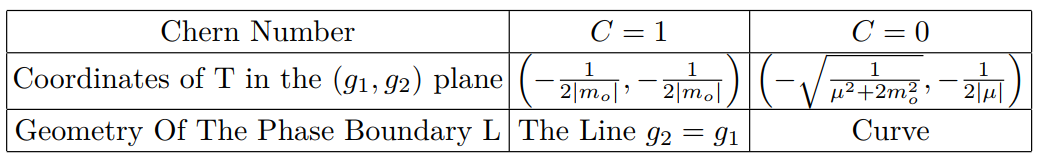}
  \caption*{}
  \label{tables}
\end{figure}

\section{Discussion and conclusions}
We leverage the van der Waerden notation, widely used in supersymmetric spinor calculations, to map the low-energy effective Hamiltonian of the Haldane model with Hubbard interaction onto a Lorentz-invariant Lagrangian featuring four-fermion interactions. 
By solving this model, we unveil two distinct phases: superconducting phase and CDW phase, both driven by the Hubbard interactions.
Through this mapping, we establish a connection between high-energy and condensed matter physics, where bilinear quantities and four-Fermi terms in the Lorentz-invariant Lagrangian can be derived in real condensed matter systems, facilitating the emulation of high-energy phenomena in condensed matter systems.
For example, the term $\bar{\Psi}_i \mu \gamma^3 \Psi_i$ originates from the chemical potential added to sublattices A and B in the honeycomb lattice, with $\pm \mu$, and $\bar{\Psi}_i m_o \gamma^3 \gamma^5 \Psi_i$ ($m_o = 3 \sqrt{3} t^{\prime} \sin \phi$) arises from next-nearest-neighbor interactions and the  magnetic flux. The interaction term:
$H_{\text{int}} = \frac{U}{4} \left[ (\Psi^{\intercal} C \Psi)(\bar{\Psi} C \bar{\Psi}^{\intercal}) + (\bar{\Psi}\Psi)^2 - (\bar{\Psi}i\gamma^5\Psi)^2 \right]$
stems from the Hubbard interaction. These mappings substantiate the physical significance of our Lagrangian and its parameters.
Our approach is novel in that it provides a methodology to simulate phenomena of interest in high-energy physics within condensed matter systems, thereby fostering a two-way exchange of ideas and techniques between these fields. Such cross-fertilization enriches both domains, offering new insights into fundamental physical processes.
Looking forward, there are several promising avenues for future research. 
Investigating other two-dimensional lattice models with topological properties \cite{ssun2013topological,sun2023topological1,sun2023topological2,kklipstein2021hard,cchen2015tunable,zzhao2021coulomb,li2020topological,li2017topological,lee2014lattice,zhang2014honeycomb,yang2019robust} and exploring their connection to high-energy physics through similar mapping techniques would provide a broader understanding of the interplay between topology and phase transitions in various systems. 
Extending our approach to systems with more complex interactions such as long-range couplings could shed light on the emergence of exotic phases and help identify novel materials with unique properties for potential applications in quantum information and nanotechnology.

\section{Acknowledgements}
This work was supported by National Key R\&D Program of China under grants No. 2021YFA1400900, 2021YFA0718300, 2021YFA1402100, NSFC under grants Nos. 61835013, 12174461, 12234012, Space Application System of China Manned Space Program.

\appendix

\section{Algebra of the $\gamma$ matrices } \label{aa}
Given that the two spinors in graphene, which are expanded at two inequivalent Dirac points, 
correspond to left and right-handed fermions respectively, we adopt the so-called Weyl representation for our gamma algebra here \cite{peskin2018introduction,srednicki2007quantum}:
\begin{equation}
  \gamma^{\mu}=\left(\begin{array}{ll}
  0 & \bar{\sigma}^{\mu} \\
  \sigma^{\mu} & 0
  \end{array}\right),
  \end{equation}
where $\mu=0,1,2,3$ and $\bar{\sigma}^{\mu}=\sigma^{\mu}$ when $\mu=0$ and $\bar{\sigma}^{\mu}=-\sigma^{\mu}$ when $\mu=1,2,3$ (where $\sigma_0$ is the unit 2 $\times$ 2 matrix $I$). These gamma matrix have the properties:
$\{\gamma^\mu,\gamma^\nu\}=2g^{\mu \nu}$,where $g^{\mu \nu}=g_{\mu \nu}=diag(1,-1,-1,-1)$. There exist another matrices $\gamma^5$, which anticommute with all $\gamma^\mu$:
\begin{equation}
\gamma^5=i \gamma^0 \gamma^1 \gamma^2 \gamma^3=\left(\begin{array}{ll}
  -I & 0 \\
 0 & I
  \end{array}\right).
\end{equation}
Trace of $\gamma$ matrices can be evaluated as follows:
\begin{equation}
  \begin{aligned}
  \operatorname{tr}(\mathbf{1}) & =4, \\
  \operatorname{tr}\left(\text { any odd  } \gamma^{\prime} \mathrm{s}\right) & =0, \\
  \operatorname{tr}\left(\gamma^\mu \gamma^\nu\right) & =4 g^{\mu \nu}, \\
  \operatorname{tr}\left(\gamma^\mu \gamma^\nu \gamma^\rho \gamma^\sigma\right) & =4\left(g^{\mu \nu} g^{\rho \sigma}-g^{\mu \rho} g^{\nu \sigma}+g^{\mu \sigma} g^{\nu \rho}\right), \\
  \operatorname{tr}\left(\gamma^5\right) & =0, \\
  \operatorname{tr}\left(\gamma^\mu \gamma^\nu \gamma^5\right) & =0, \\
  \operatorname{tr}\left(\gamma^\mu \gamma^\nu \gamma^\rho \gamma^\sigma \gamma^5\right) & =-4 i \epsilon^{\mu \nu \rho \sigma}.
  \end{aligned}
  \end{equation}
Contractions of $\gamma$ matrices with each other simplify to:
\begin{equation}
  \begin{aligned}
  \gamma^\mu \gamma_\mu & =4, \\
  \gamma^\mu \gamma^\nu \gamma_\mu & =-2 \gamma^\nu, \\
  \gamma^\mu \gamma^\nu \gamma^\rho \gamma_\mu & =4 g^{\nu \rho}, \\
  \gamma^\mu \gamma^\nu \gamma^\rho \gamma^\sigma \gamma_\mu & =-2 \gamma^\sigma \gamma^\rho \gamma^\nu.
  \end{aligned}
  \end{equation}

\section{Performing the path integral over the fermion} \label{bb}
Let us show here some of the details of the path integral
over the fermions  and which leads to the effective thermodynamic potential. Adopting the procedure
described in Refs.~ \cite{core,klimenko2012superconducting}, we assume two anti-commuting
four component Dirac spinor fields $q(x)$ and $\bar{q}(x)$. Then,we'll calculate the following path integral:
\begin{eqnarray}
    I &=& \int Dq D\bar{q}~ e^{ i \int d^3x \left[ \bar{q} \mathcal{D} q -
        \frac{\Delta^*}{2} q^\intercal C q - \frac{\Delta}{2} \bar{q} C \bar{q}^\intercal
        \right] },
    \end{eqnarray}
   
    where $\mathcal{D} = i  \gamma_\mu \partial_\nu + \mu
    \gamma^3+m_o \gamma^3 \gamma^5- \sigma-i\pi \gamma^5$ and $C =- i \gamma^2 \gamma^0$ is the charge conjugation
    matrix. Using the Gaussian path integral identities
  
    \begin{equation}
      \int Dp e^{ i \int d^3x \Big[ -\frac{1}{2} p^\intercal A p + \eta^\intercal p \Big]} = (\det A)^{1/2} e^{ -\frac{i}{2} \int d^3x \eta^\intercal A^{-1} \eta }~,
    \end{equation}

    and 
    \begin{equation}
      \int D\bar{p} e^{ i \int d^3x \Big[ -\frac{1}{2} \bar{p} A \bar{p}^\intercal + \eta \bar{p}^\intercal \Big] } = (\det A)^{1/2} e^{ -\frac{i}{2} \int d^3x \bar{\eta} A^{-1} \bar{\eta}^\intercal },
      \end{equation}

    and by also considering $A = \Delta C$, $\bar{q} \mathcal{D} =
    \eta^\intercal$, $\mathcal{D}^\intercal \bar{q}^\intercal = \eta$, one finds, after
    integrating over $q$ and $\bar{q}$, the result

    \begin{align}
      I &= \int Dq D\bar{q}~ e^{ i \int d^3x \Bigg[ \bar{q} \mathcal{D} q -
        \frac{\Delta^*}{2} q^\intercal C q - \frac{\Delta}{2} \bar{q} C \bar{q}^\intercal
        \Bigg] }  \nonumber \\
       &= \left[\det \left( \Delta^2 C^2+ \mathcal{D} C^{-1}
      \mathcal{D}^\intercal C \right) \right]^{\frac{1}{2}}, 
      \end{align}

    where we have assumed $\Delta = \Delta^*$ in the last step . Using the relations $C^{-1}
    \gamma_\mu^\intercal C = - \gamma_\mu$ ($\mu =0,1,2,3$) and $\partial_\mu^\intercal = - \partial_\mu$,
    one finds that
    \begin{equation}
     I = [\det(-\Delta^2 + \mathcal{D}_+ \mathcal{D}_-)]^{1/2} = (\det
     B)^{\frac{1}{2}},
    \end{equation}
    with $\mathcal{D}_\pm = i  \gamma_\mu \partial_\nu +m_o \gamma^3 \gamma^5- \sigma-i\pi \gamma^5 \pm \mu
    \gamma^3$. {}Finally, using the identity $\det B = \exp({\rm
      Tr } \ln B)$ one finds
    \begin{equation} \label{lni}
     \ln I = \frac{1}{2} {\rm Tr} ( \ln B )= \int d^3x  \sum_{i} \int
     \frac{d^3p}{(2 \pi)^3} \ln \lambda_i(p),
    \end{equation}  
    where $\lambda_i(p)$ are the eigenvalues of matrix $B(p)$:
    \begin{align}
        B(p)&=-\Delta^2+(p_{\mu}\gamma^{\mu}+m_o \gamma^3 \gamma^5-\sigma-i\gamma^5 \pi+\mu \gamma^3) \nonumber \\
            &\times(p_{\mu}\gamma^{\mu}+m_o \gamma^3 \gamma^5-\sigma-i\gamma^5 \pi-\mu \gamma^3) \nonumber \\
            &=p^2-\Delta^2+m_o^2+\sigma^2+\mu^2 \nonumber \\
            &-2\sigma \slashed{p}-2\sigma m_o \gamma^3 \gamma^5+2m_o \slashed{p} \gamma^3 \gamma^5  -2\mu \slashed{p}\gamma^3 -2m_o \mu \gamma^5,
\end{align}

and $\slashed{p}=p^\mu \gamma_\mu$. We first consider the integration over the frequency component $p_0$. The eigenvalues $\lambda_i$ of the matrix $B(p)$ can be expressed as a polynomial in terms of $p_0$ as follows:
\begin{equation}
  \lambda_i(p)=p_0-\widetilde{\lambda_i}({\bf p}),
\end{equation}
where
\begin{equation}
    \widetilde{\lambda_i}({\bf p})=\left\{\begin{array}{l}
    \pm \sqrt{{\bf p}^2+(\Delta+m_o+\beta \sqrt{\mu^2+\sigma^2})^2},\\
    \pm \sqrt{{\bf p}^2+(\Delta-m_o-\beta \sqrt{\mu^2+\sigma^2})^2},
    \end{array}\right.
\end{equation}
and $\beta=\pm$. This implies that $p_0$ has a total of eight roots. 

To integrate the frequency part of Eq.~(\ref{lni}), we employ the formula $\int_{-\infty}^\infty dp_0\ln\big(p_0-R)=\mathrm{i}\pi|R|$, yielding:
\begin{align}
  \ln I &= i \int d^3 x \sum_{\beta} \int \frac{d^2 {\bf p}}{(2\pi)^2} \notag\\  
  &\Bigg( \Big| \sqrt{{\bf p}^2+(\Delta+m_o+\beta \sqrt{\mu^2+\sigma^2})^2} \Big| \nonumber \\ 
  &\quad + \Big| \sqrt{{\bf p}^2+(\Delta-m_o-\beta \sqrt{\mu^2+\sigma^2})^2} \Big| \Bigg).
\end{align}

From this, we can derive the unrenormalized thermodynamic potential (TDP) as:
\begin{equation}
  \Omega \equiv \Omega^{un}(\Delta,\sigma)=\frac{\sigma^2}{4G}+\frac{\Delta^2}{4G} -\frac{1}{4} \sum_{\beta,i} \int \frac{d^2 {\bf p}}{(2\pi)^2} |p^{\beta} _{0i}|,
\end{equation}
where $p_{0,1} ^\beta=\sqrt{{\bf p}^2+(\Delta+m_o+\beta \sqrt{\mu^2+\sigma^2})^2}$ and $p_{0,2} ^\beta=\sqrt{{\bf p}^2+(\Delta-m_o-\beta \sqrt{\mu^2+\sigma^2})^2}$.

To renormalize the TDP, we use the following formula:
\begin{align}
  &\Omega^{un} (\Delta,\sigma,\mu,m_o) = \frac{\sigma^2}{4G}+\frac{\Delta^2}{4G} \notag \\
  &\quad -\int \frac{d^2 {\bf p}}{(2\pi)^2} \Bigg( \sqrt{{\bf p}^2+(\Delta+\sigma)^2}+\sqrt{{\bf p}^2+(\Delta-\sigma)^2} \Bigg) \notag \\
  &\quad - \frac{1}{4} \sum_{\beta,i} \int \frac{d^2 {\bf p}}{(2\pi)^2} \Bigg( |p^{\beta} _{0i}| \notag \\
  &\quad - 4 \Bigg( \sqrt{{\bf p}^2+(\Delta+\sigma)^2}+\sqrt{{\bf p}^2+(\Delta-\sigma)^2} \Bigg) \Bigg) \notag \\
  &= V^{un}(\Delta,\sigma) \quad - \frac{1}{4} \sum_{\beta,i} \int \frac{d^2 {\bf p}}{(2\pi)^2} \Bigg( |p^{\beta} _{0i}| \notag \\
  &\quad - 4 \Bigg( \sqrt{{\bf p}^2+(\Delta+\sigma)^2}+\sqrt{{\bf p}^2+(\Delta-\sigma)^2} \Bigg) \Bigg).
\end{align}

The second term on the right-hand side of the above equation is finite, so the renormalization of $\Omega$ depends only on the renormalization of the first term. As we have derived in the main text, the renormalized potential is given by $ V^{ren}(\Delta,\sigma)=\frac{\sigma^2}{2 \pi g}+\frac{\Delta^2}{2 \pi g} 
  +\frac{1}{6\pi}(|\Delta+\sigma|^3+|\Delta-\sigma|^3)$. 

Finally, we can express the renormalized TDP as:
     \begin{align}   \label{RTDP}
        &\Omega^{ren}(\Delta,\sigma) = \frac{\sigma^2}{2 \pi g_1} + \frac{\Delta^2}{2 \pi g_2}  \notag \\
        &+\frac{1}{12\pi} (|\Delta+m_o+\sqrt{\mu^2+\sigma^2}|^3 + |\Delta+m_o-\sqrt{\mu^2+\sigma^2}|^3  \notag \\
        &+|\Delta-m_o+\sqrt{\mu^2+\sigma^2}|^3 + |\Delta-m_o-\sqrt{\mu^2+\sigma^2}|^3)   \,.
    \end{align}

\section{ Using van der Waerden notation to construct Lorentz invariants}\label{cc}
    In this section we show some of the details for the derivation of bilinear and four fermion terms.
    %
In the main text, we let  $\Psi=\left(\begin{array}{l}\eta \\ \chi\end{array}\right)$. where $\eta=\left(\begin{array}{l} a_+ \\ b_+ \end{array}\right)$, $\chi=\left(\begin{array}{l}b_- \\ a_-\end{array}\right)$ have the properties of right-chiral and left-chiral weyl spinor respectively.
So, according to the van der Waerden notation, we use lowwer undotted indices to denote the component of the left-chiral weyl spinor, and use upper dotted indices with a bar symbol over spinors to denote the components of right-chiral spinors. And we use the Levi-Civita symbol $\epsilon^{ij}, \; \epsilon_{ij}$  to raise or lower both dotted and undotted indices \cite{muller2010introduction,labelle2010supersymmetry,martin2010supersymmetry}.
\begin{equation}
    \begin{split}
     &\chi_1=b_-, \; \chi_2=a_-,\;\chi^a=\epsilon^{ab}\chi_b, \\
     &\bar{\eta}^{\dot{1}}=a_+  , \;\bar{\eta}^{\dot{2}}=b_+ ,\;\bar{\eta}^{\dot{a}}=\epsilon^{\dot{a}\dot{b}}\bar{\eta}_{\dot{b}},\\
     &\bar{\chi}_{\dot{a}}=\chi_a ^\dagger, \; \bar{\chi}^{\dot{a}}=\chi{^a} ^{\dagger},\; \bar{\eta}_{\dot{a}}=\eta_a ^\dagger,\; \bar{\eta}^{\dot{a}}=\eta{^a} {^\dagger}, \\
     &\epsilon^{12}=-\epsilon_{12}=1.
    \end{split}
    \end{equation}
    Especially, for the right-chiral spinor $\eta$, we use $\bar{\eta}$ to denote that we will employ the dot/undot indices, and we use $\eta^\dagger$ to signify that we will utilize the true components of the Hermitian conjugate of $\eta$, i.e. $\eta^\dagger_1=a_{+} ^\dagger$ and $\bar{\eta}_{\dot{1}}=\epsilon_{\dot{1}\dot{2}}\bar{\eta}^{\dot{2}}=-b_+$.
 First, let's deal with the efective chemical potential,
 the effective chemical potential is:
\begin{equation}
    H_\mu=\mu_A (a_+ ^\dagger a_+ +a_- ^\dagger a_-) + \mu_B  (b_+ ^\dagger b_+ +b_- ^\dagger b_-).
\end{equation}
Using the van der waerden notation, the first term can be written as:
\begin{align}
  \mu_A a_+ ^\dagger a_+&=\frac{\mu_A }{2} (\eta^1 \sigma^3 _{1\dot{1}} \bar{\eta}^{\dot{1}}+\eta^1 \sigma^0 _{1\dot{1}} \bar{\eta}^{\dot{1}}+\eta^2 \sigma^3 _{2\dot{2}} \bar{\eta}^{\dot{2}}+\eta^2 \sigma^0 _{2\dot{2}} \bar{\eta}^{\dot{2}}) \notag \\
  &=\frac{\mu_A}{2}(\eta \sigma^3 \bar{\eta} + \eta \sigma^0 \bar{\eta}).
\end{align}
Similarly, the remaining three can be written as:
\begin{equation}
  \begin{split}
    \mu_A a_- ^\dagger a_-&=\frac{\mu_A}{2} (\bar{\chi}\bar{\sigma}^3\chi + \bar{\chi} \bar{\sigma}^0 \chi), \\
    \mu_B b_+ ^\dagger b_+ &= -\frac{\mu_B}{2} (\eta \sigma^3 \bar{\eta} - \eta \sigma^0 \bar{\eta}), \\
    \mu_B b_- ^\dagger b_- &= -\frac{\mu_B}{2} (\bar{\chi}\bar{\sigma}^3\chi - \bar{\chi} \bar{\sigma}^0 \chi).
  \end{split}
\end{equation}
Hence, we obtain:
\begin{equation}
  H_\mu=\frac{\mu_A+\mu_B}{2}(\eta \sigma^0 \bar{\eta}+\bar{\chi} \bar{\sigma}^0 \chi) +\frac{\mu_A-\mu_B}{2}(\eta \sigma^3 \bar{\eta}+\bar{\chi} \bar{\sigma}^3 \chi).
\end{equation}
We note that the components of the dirac spinors actually is: $\Psi=\left(\begin{array}{l}\bar{\eta}^{\dot{a}} \\ \chi_a\end{array}\right)$, and $\bar{\Psi}=(\bar{\chi}_{\dot{a}},\eta^a)$.
 So we have: $\eta \sigma^{\mu} \bar{\eta}+\bar{\chi} \bar{\sigma}^{\mu} \chi=\bar{\Psi} \gamma^{\mu} \Psi$.
Final, the effective chemical potential can be written as :
\begin{equation}
  H_\mu=\frac{\mu_A+\mu_B}{2} \bar{\Psi} \gamma^{0} \Psi+ \frac{\mu_A-\mu_B}{2} \bar{\Psi} \gamma^{3} \Psi.
\end{equation}
Second, we deal with the Haldane mass term: $\psi_{\pm}^{\dagger} m_{\pm} \sigma_{z} \psi_{\pm}$, the sum of these two terms can be written as:
\begin{align}  
  H_o&=\psi_{+}^{\dagger} m_{o} \sigma_{z} \psi_{+}+\psi_{-}^{\dagger} (-m_{o}) \sigma_{z} \psi_{-}\notag \\
  &=m_o(\eta^a \sigma^3 _{a \dot{b}} \bar{\eta}^{\dot{b}} -\bar{\chi}_{\dot{a}} \bar{\sigma}^{3 \dot{a} b} \chi_b) \notag \\
  &=m_o \bar{\Psi} \gamma^3 \gamma ^5 \Psi.
\end{align}

Finally, we will derive the four-fermion interactions which are lorentz invariant.
In the main text, we obtain 
\begin{align}
  H_{int} &\approx \int d {\bf r} \ U (a_{+} ^{\dagger}a_{+} b_{+} ^{\dagger}b_{+}+a_{-} ^{\dagger}a_{-}b_{-} ^{\dagger}b_{-}) \notag \\
  &=\int d {\bf r} \ \frac{U}{4}  [( \eta \cdot \eta)(\bar{\eta} \cdot \bar{\eta})+(\bar{\chi} \cdot \bar{\chi})(\chi \cdot \chi)].
\end{align}
We note $\Psi^{\intercal} C \Psi=(\eta^{\intercal},\chi^{\intercal}) \left(\begin{array}{cc}
  -i\sigma^2 & 0 \\
  0 & i\sigma^2
  \end{array}\right) \left(\begin{array}{l}\eta \\ \chi\end{array}\right) = \bar{\eta}\cdot \bar{\eta}+\chi \cdot \chi$ and $\bar{\Psi} \Psi=(\bar{\chi}_{\dot{a}},\eta^a)\left(\begin{array}{l}\bar{\eta}^{\dot{a}} \\ \chi_a\end{array}\right)=\bar{\chi}\cdot\bar{\eta}+\eta \cdot \chi$.
  Therefore, we can get:
  \begin{equation}
    \begin{split}
    &( \eta \cdot \eta)(\bar{\eta} \cdot \bar{\eta})+(\bar{\chi} \cdot \bar{\chi})(\chi \cdot \chi) \\
    &=(\bar{\eta}\cdot\bar{\eta}+\chi \cdot \chi)(\bar{\chi}\cdot\bar{\chi}+\eta \cdot \eta)-\bar{\eta}\cdot\bar{\eta}\bar{\chi}\cdot\bar{\chi}-\chi \cdot \chi \eta \cdot \eta \\
    &=(\Psi^{\intercal} C \Psi)(\bar{\Psi} C \bar{\Psi}^{\intercal})+2(\eta \cdot \chi)^2+2(\bar{\eta}\cdot\bar{\chi})^2 \\
  &=(\Psi^{\intercal} C \Psi)(\bar{\Psi} C \bar{\Psi}^{\intercal})+(\eta \cdot \chi +\bar{\eta}\cdot \bar{\chi})^2+(\eta \cdot \chi -\bar{\eta}\cdot \bar{\chi})^2  \\
  &=(\Psi^{\intercal} C \Psi)(\bar{\Psi} C \bar{\Psi}^{\intercal})+(\bar{\Psi}\Psi)^2-(\bar{\Psi}i\gamma^5\Psi)^2.
\end{split}
  \end{equation}
  
By the way, if we consider the full effective Hubbard interactions: $\mathcal{H}_F= \int d {\bf r}\  U (a_{+} ^{\dagger}a_{+} b_{+} ^{\dagger}b_{+}+a_{-} ^{\dagger}a_{-}b_{-} ^{\dagger}b_{-}+a_{+} ^{\dagger}a_{+}b_{-} ^{\dagger}b_{-}+a_{-} ^{\dagger}a_{-}b_{+} ^{\dagger}b_{+})$, we can map this into :
\begin{equation}
  \mathcal{H}=(\bar{\Psi} \gamma^0 \Psi)^2-(\bar{\Psi} \gamma^3 \Psi)^2.
\end{equation}
Finally, we present the expressions for the charge-density-wave (CDW) and superconducting (SC) order parameters as derived in the main text:
\begin{equation}
  \begin{split}
    \langle\sigma \rangle&=-2 \frac{G}{N} \langle \bar{\Psi} \Psi \rangle =-2 \frac{G}{N} \langle a_{+} ^{\dagger}b_{-}+a_{-} ^{\dagger}b_{+}+b_{+} ^{\dagger}a_{-}+b_{-} ^{\dagger}a_{+}\rangle,\\
    \langle \Delta \rangle&=-2 \frac{G}{N} \langle \Psi ^\intercal C \Psi \rangle = -4 \frac{G}{N} \langle a_{+} b_{+} + a_{-} b_{-}\rangle.
  \end{split}
\end{equation}

\section{$(g,\mu),(g,m_o)$ phase diagrams when $m_o\neq0$ and $\mu \neq 0$}\label{dd} \label{appen4}
This section illustrates the phase diagrams of the system in the $(g,\mu)$ and $(g,m_o)$ planes, given non-zero values of $m_o$ and $\mu$, respectively. The $(g,m_o)$ phase diagram reveals that even for infinitesimally small non-zero chemical potential $\mu$, the system transitions from the initial MIX phase (as described in the main text) to the CDW phase. This observation underscores the pivotal role of $\mu$ in driving the system towards the CDW phase, while concurrently suppressing the superconducting phase to a certain extent.
\begin{figure*}[htbp]
	\centering
	\includegraphics[width=0.9\textwidth]{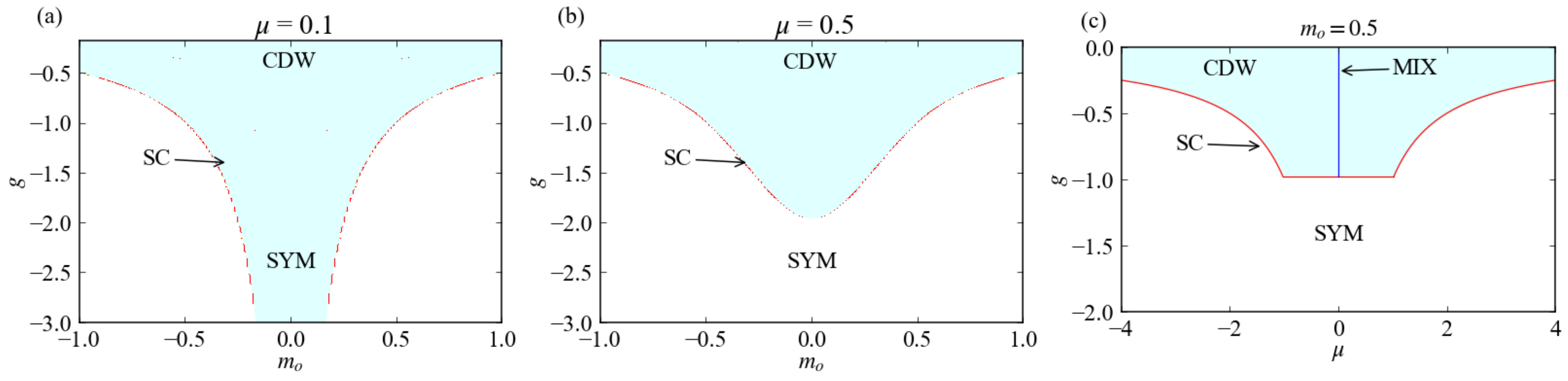}
  \caption{Phase diagrams in the $(g,\mu)$ and $(g,m_o)$ planes for non-zero $m_o$ and $\mu$.
(a) and (b) depict the $(g,m_o)$ phase diagrams for $\mu=0.1$ and $\mu=0.5$, respectively. (c) illustrates the $(g,\mu)$ phase diagram for $m_o=0.5$.}
 \label{fig-S1}
 \end{figure*}

\section{Realization proposal using ultracold atoms} \label{experiments}
In a realistic experiment, we note that the realization of the Haldane-Hubbard model is already attainable in cold atoms, and it is possible now to study the quantum phase transitions discussed here.
To be specific, we can use a polarized ultracold Fermi gas of $^{40}$K atoms, here we choose the hyperfine state
$|F,m_F\rangle=|9/2,-7/2\rangle$ by applying certain magnetic fields
\cite{Tarruell2012nat,Jotzu2014nat},
and tunable $p$-wave interactions have been well implemented utilizing this species of atom in experiments
\cite{Regal2003prl,Gunter2005prl,Ticknor2004pra}.
We load the atoms into the honeycomb optical lattice at a wavelength of $\lambda=826$ nm with potential
$V(x,y)=-V_{x1}\sin^2(k_Lx)-V_{x2}\cos^2(k_Lx)-V_y\cos^2(k_Ly)-2\sqrt{V_{x2}V_y}\cos(k_x)\cos(k_y)$.
Here $V_{x1}=5.0E_R,V_{x2}=0.46E_R$ are the lattice depths of two collinear beams along $x$ to create a standing wave,
$V_y=2.2E_R$ is the lattice depth of the perpendicular beam along $y$ to create a spacing honeycomb lattice,
and $k_L=2\pi/\lambda$.
The recoil energy of such an optical lattice is $E_R=h^2/(2m\lambda^2)\approx 7.26$kHz.
The frequency detuning between the $x_1$ and $x_2$ beams should be set as $\pi$, whereas the same frequency for $x_2$ and $y$ beams, to control the energy offset $\mu$ between A-B sublattices.
And the phase difference of beams along $x$ and $y$ can control the imaginary part of NNN hopping.
Using a Wannier function calculation
\cite{Uehlinger2013prl}, 
we extract the amplitude of the NN hopping $t\approx 530$Hz,
and the NNN hopping $t'\approx 15$Hz.

Using the technique of $p$-wave Feshbach resonance, we can obtain Hubbard interaction $U$ as mentioned.
The quantum phase transitions here are signaled by the several quasiparticle gaps,
which would have dramatic effects on dynamic structure factor that can be measured by Bragg spectroscopy
\cite{Vogels2002prl,Clement2009prl,Ernst2010natphys}.
We expect that, the charge-density wave gap $\sigma$ and superconductor gap $\Delta$  can be distinguished by different characteristics by spectroscopy measurement.
\section{Higher-order corrections to the effective potential.} \label{high}
To illustrate the efficacy of the mean-field approximation within the large N framework, we will calculate higher-order corrections to the effective potential.
This approach follows the methodology established in \cite{marino2006quantum}, which have meticulously explored quantum fluctuations under similar conditions.
We aim to prove that quantum fluctuations at zero temperature do not invalidate the saddle-point solutions derived from the mean-field calculations. Specifically, we need to show that the symmetry phase (\(\Delta=0\), \(\sigma=0\)) is not disrupted by quantum fluctuations. For clarity, we will discuss this using the CDW phase as an example.

The Lagrangian we consider is given by:
\begin{equation}
    \mathcal{L}=\bar{\Psi}_i \left(i \gamma^{v} \partial_{v} +\mu \gamma^3 + m_o \gamma^3 \gamma^5 -\sigma \right) \Psi_i -\frac{N}{4G}\sigma^2.
\end{equation}

Upon integrating out the fermionic fields, we obtain the effective action:
\begin{equation}
\mathcal{S}_{\text{eff}} [\sigma]=\int d^3 x \left(-\frac{N}{4G} \sigma^2\right)-iN \ln \text{Det}[\mathcal{D}[\sigma]]
\end{equation}
where
\(\text{Det}[\mathcal{D}[\sigma]]=[\sigma^2+\mu^2-m_o ^2-p^2]^2-4m_o ^2 p^2\), and \(p^2=p_0 ^2-p_1^2-p_2 ^2\).

Correspondingly, the effective potential is:
\begin{align} \label{ef}
    \Omega_{\text{eff}}(\sigma)&=\frac{1}{4G} \sigma^2 \notag \\
    &+i\int \frac{d^3p}{(2\pi)^3} \log\left([\sigma^2+\mu^2-m_o ^2-p^2]^2-4m_o ^2 p^2\right).
\end{align}

Our objective is to demonstrate that the symmetry phase is robust against fluctuations. As \(\mu\) and \(m_o\) serve primarily as parameters to control the phase boundaries, for simplicity, we set them to zero and discuss in Euclidean space. The effective potential then becomes:
\begin{equation}
    \Omega_{\text{eff}}(\sigma)=\frac{1}{4G} \sigma^2-2\int \frac{d^2k}{(2\pi)^2}\int \frac{d \omega}{2\pi}\log(\omega ^2+k^2+\sigma^2).
\end{equation}

Now let's think about fluctuations. The fluctuations are represented as \(\sigma \rightarrow \sigma + \eta\). Integrating out these fluctuations \(\eta\), we obtain the modified effective action:
\begin{equation}
    \tilde{S}_{\text{eff}}[\sigma] = S_{\text{eff}}[\sigma] - \text{Tr} \ln \mathcal{A}[\sigma],
\end{equation}
where \(\mathcal{A}[\sigma]\) is defined as:
\begin{equation}
    \mathcal{A}[\sigma] = 
\begin{pmatrix}
\frac{\delta^2 S}{\delta \sigma \delta \sigma^*} & \frac{\delta^2 S}{\delta \sigma^2} \\
\frac{\delta^2 S}{\delta \sigma^{*2}} & \frac{\delta^2 S}{\delta \sigma^* \delta \sigma} 
\end{pmatrix}
[\sigma].
\end{equation}

The corresponding fluctuation-considered effective potential is given by:
\begin{equation}
    \tilde{\Omega}_{\text{eff}}(\sigma) = \Omega_{\text{eff}}(\sigma) + \Omega_{f}(\sigma),
\end{equation}
where \(\Omega_{\text{eff}}(\sigma)\) is provided by Eq.~(\ref{ef}), and \(\Omega_{f}(\sigma)\) is expressed as:
\begin{align}
  \Omega_{f}(\sigma) &= - \int \frac{d^2k}{(2\pi)^2} \int \frac{d\omega}{2\pi} \left[ \ln \left( \frac{1}{4G} - \alpha(\omega, k) \right) \right. \notag \\ 
  &\left. + \ln \left( \frac{1}{4G} - \alpha(\omega, k) + 2\sigma^2 \beta(\omega, k) \right) \right],
\end{align}
with \(\alpha(\omega, k, \sigma)\) and \(\beta(\omega, k, \sigma)=-\frac{\partial \alpha}{\partial \sigma^2}\) defined as:
\begin{equation}
    \alpha(\omega, k, \sigma) = \int \frac{d^2q}{(2\pi)^2} \int \frac{d\theta}{2\pi} \frac{2}{[ |q| +i\theta] [|q+k|- i(\theta + \omega)] + \sigma^2 }.
\end{equation}
It is observed that:
\begin{equation}
    \Omega_{f}^{\prime}=|\sigma| V(|\sigma|),
\end{equation}
with \(V(0)\) being a finite constant, indicating that \(\sigma=0\) is a solution to the gap equation \(\tilde{\Omega}_{\text{eff}}^{\prime}(\sigma)=0\) with fluctuations considered. The second derivative of the effective potential, \(\tilde{\Omega}_{\text{eff}}^{\prime \prime}(\sigma)\), is given by:
\begin{equation}
    \tilde{\Omega}_{\text{eff}}^{\prime \prime}(\sigma) = \Omega_{\text{eff}}^{\prime \prime}(\sigma) + \Omega_{f}^{\prime \prime}(\sigma),
\end{equation}
Notably, in our calculations for convenience, we set \(\hbar=1\). However, in reality, \(\frac{\Omega_{f}^{\prime \prime}}{\Omega_{\text{eff}}^{\prime \prime}}\propto \hbar^2\). Thus, \(\Omega_{f}^{\prime \prime}(\sigma=0)\) is significantly smaller than \(\Omega_{\text{eff}}^{\prime \prime}(\sigma=0)\), implying that \(\tilde{\Omega}_{\text{eff}}^{\prime \prime}(\sigma=0)\) and \(\Omega_{\text{eff}}^{\prime \prime}(\sigma=0)\) have the same sign. Therefore, even considering fluctuations, \(\sigma=0\) is indeed a minimum. Similarly, the charge density wave (CDW) solution \(\sigma \neq 0\) in the mean-field approach is not nullified by quantum fluctuations, indicating that our zero-temperature phase diagram derived from the mean-field is robust against fluctuations.

\bibliographystyle{apsrev4-1}
\bibliography{refer}


\end{document}